\documentclass[11pt]{article}
\usepackage[english]{babel}
\usepackage{hyperref,amsthm,amsmath,amssymb,color,multirow,graphicx }
\usepackage{inputenc}
\usepackage{wrapfig, float,array}

\newtheorem{remark}{Remark}[section]

\newtheorem*{thm*}{�������}
\newtheorem{lem}{Lemma}[section]
\theoremstyle{definition}

\numberwithin{equation}{section}

\begin{document}

\begin{center}
\textbf{\large Ground states for the SOS model with an external field on the Cayley tree }\\
\textbf{ M.M.Rahmatullaev, M.R.Abdusalomova, M.A.Rasulova}
\end{center}

{\small \textbf{Abstract.} We consider a nearest-neighbor
solid-on-solid (SOS) model, with several spin values $0,1,2,...,m,
m\geq2$  and non zero external field, on a Cayley tree of order
$k$. In the case $k=2, m=2$, we describe translation-invariant
ground states for the SOS model with a translation-invariant
external field. Some periodic ground states for the SOS model with
periodic external field are described.

\textbf{Keywords:}  Cayley tree, SOS model, external field,
translation-invariant external field, periodic external field,
configuration, translation-invariant ground state, periodic ground
state.

\makeatletter
\renewcommand{\@evenhead}{\vbox{\thepage \hfil {\it M.M.Rahmatullaev, M.R.Abdusalomova, M.A.Rasulova}   \hrule }}
\renewcommand{\@oddhead}{\vbox{\hfill
{\it Ground states for the SOS model with an external field on the Cayley tree }\hfill \thepage
\hrule}} \makeatother

\label{firstpage}

\section{Introduction} \label{sec:intro}

One of fundamental problems is to describe the extreme Gibbs
measures corresponding to a given Hamiltonian. Each Gibbs measure
is associated with a single phase of a physical system. Existence
of two or more Gibbs measures means that phase transitions exist.

As is known, the phase diagram of Gibbs measures for a Hamiltonian
is close to the phase diagram of isolated (stable) ground states
of this Hamiltonian.At low temperatures, a periodic ground state
corresponds to a periodic Gibbs measure.Therefore the problem of
description of periodic ground states naturally arises (see [1],
[3],[5]-[9]).

We consider a nearest-neighbor solid-on-solid (SOS) model, with
several spin values $0,1,2,...,m, m\geq2$  and non zero external
field, on a Cayley tree of order $k$. The SOS model can be treated
as a natural generalization of the Ising model (obtained for
$m=1$). We mainly assume that $m=2$ (three spin values) and study
translation-invariant and periodic ground states.

In [5] and [6],[8] for the Ising model with competing interactions, periodic and weakly periodic
ground states were described.

In [7] for the Potts model with competing interactions on the Cayley tree
of order $k$ with $k\geq 2$, periodic and weakly periodic ground
states for normal subgroups of index 4 were studied.

In [3] for the $\lambda-$model on the Cayley tree of order two, periodic
and weakly periodic ground  states were studied.

In [9] for the Ising model on the Cayley tree of order two, translation-invariant, periodic ground states were described.

In this paper we shall study translation-invariant and periodic ground states for the SOS model with external fields.

\section{Main definitions and known facts}

Let $\tau^k = (V,L)$ be a Cayley tree of order
$k$, i.e, an infinite tree such that exactly $k+1$ edges are incident to each vertex. Here $V$ is
the set of vertices and $L$ is the set of edges of $\tau^k$.

Let $G_k$ denote the free product of $k+1$ cyclic groups $\{e, a_i\}$ of order 2 with generators $a_1, a_2,\dots, a_{k+1}$,
i.e., let $a^2_i=e$ (see [4]).

There exists a one-to-one correspondence between the set $V$ of vertices of the Cayley tree of order $k$ and the group $G_k$, (see [1],[2]).

We show how to construct this correspondence. We choose an
arbitrary vertex $x_0 \in V  $and associate it with the identity
element $e$ of the group $G_k$. Since we may assume that the graph
under consideration is planar, we associate each neighbor of $x_0$
(i.e., $e$) with a single generator $a_i, i=1, 2,\dots, k + 1$,
where the order corresponds to the positive direction, see Figure
\ref{cayley}.

\begin{figure}[h!]\label{cayley}
    \begin{center}
        \includegraphics[width=12cm]{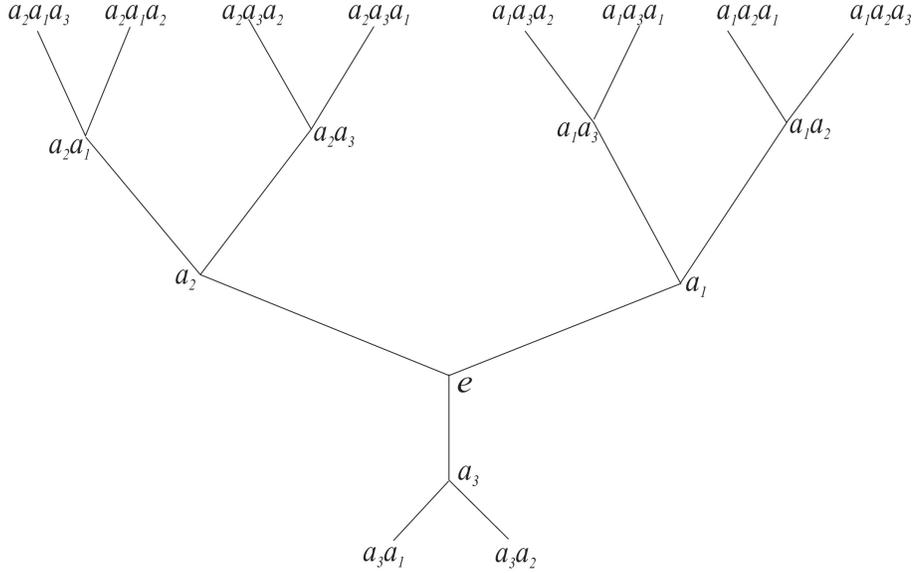}
    \end{center}
    \caption{The Cayley tree $\tau^2$ and elements of the group representation of vertices} \label{cayley}
\end{figure}

For every neighbor of $a_i$, we introduce words of the form $a_ia_j$. Since one of the neighbors of $a_i$ is $e$, we put $a_ia_i = e$. The remaining neighbors of $a_i$ are labeled according to the above order. For every neighbor of $a_ia_j$, we introduce words of length 3 in a similar way. Since one of the neighbors of $a_ia_j$ is $a_i$, we put $a_ia_ja_j = a_i$. The remaining neighbors of $a_ia_j$ are labeled by words of the form $a_ia_ja_l$, where $i, j, l = 1, 2, . . . , k + 1$, according to the above procedure. This agrees with the previous stage
because $a_ia_ja_j = a_ia^2_j= a_i$. Continuing this process, we obtain a one-to-one correspondence between the vertex set of the Cayley tree $\tau^k$ and the group $G_k$.

The representation constructed above is said to be $right$ because, for all adjacent vertices $x$ and $y$ and the corresponding elements $g,h \in G_k,$ we have either $g = ha_i$ or $h=ga_j$ for suitable $i$ and $j$. The definition of the $left$ representation is similar.

For the group $G_k$ (or the corresponding Cayley tree), we consider the left (right) shifts. For $g \in G_k$,
we put
\begin{equation*}
    T_g(h)=gh \ (T_g(h)=hg)\mbox{ \ for \ all }\ h \in G^k
\end{equation*}
The group of all left (right) shifts on $G_k$ is isomorphic to the group $G_k$.

Each transformation $S$ on the group $G_k$ induces a transformation $S$ on the vertex set $V$ of the Cayley
tree  $\tau^k$. In the sequel, we identify $V$ with $G_k$.

The following assertion is quite obvious (see \cite{1},
\cite{Gan}).\\

\textbf{Theorem 2.1.} \emph{The group of left (right) shifts on
the right (left) representation of the Cayley tree is the group of
translations.}\\

For each $x\in G_k$, let $S_1(x)$ denote the set of all neighbors
of $x$, i.e., $S_1(x)=\{y\in G_k:\langle x, y\rangle \}$, where $\langle x, y\rangle$ means that the vertex $x$ and $y$ are nearest neighbor.

Assume that spin takes its values in the set
$\Phi=\{0,1,2,...,m\}$. By a configuration $\sigma$ on $V$ we mean
a function taking $\sigma: x\in V\rightarrow\sigma(x)\in\Phi$. The
set of all configurations coincides with the set $\Omega=\Phi^V$.

Consider the quotient group $G_k/ G_k^*=\{H_1,...,H_r\}$, where
$G_k^*$ is a normal subgroup of index $r$ with $r\geq 1$.\\

\textbf{Definition 2.1.} \emph{A configuration $\sigma(x)$ is said
to be $G_k^*-$periodic if $\sigma(x)=\sigma_i$ for all $x\in G_k$
with $x\in H_i$. A $G_k-$periodic configuration is said to be
translation invariant.}\\

By \emph{period} of a periodic configuration we mean the index of
the corresponding normal subgroup.

SOS model with an external field  is given by Hamiltonian:
\begin{equation}\label{1}
H(\sigma)=-J\sum_{\langle x,y\rangle\in
L}|\sigma(x)-\sigma(y)|+\sum_{x\in V}\alpha_x\sigma(x),
\end{equation}
where $J \in \mathbb{R}$, $\alpha_{x}$ is an external field and
$\sigma\in \Omega$. As usual, $\langle x,y\rangle$ stands for
nearest neighbour vertices.

The SOS model of this type can be considered as a generalization
of the Ising model (which arises when $m=1$). Here, $J<0$ gives a
ferromagnetic (FM) and $J>0$ an antiferromagnetic (AFM) model. In
the FM case with zero external field the ground states are "flat"
configurations, with $\sigma(x)=j\in\Phi$ (there are $m+1$ of
them), in the AFM two "contrasting" checher-board configurations,
where $|\sigma(x)-\sigma(y)|=m$, $\forall <x,y>$.

Comparing with the Potts model (see, e.g., Refs. 10-14), the SOS
with zero external field has "less symmetry" and therefore more
diverse structure of phases. For example, in the FM case it is
intuitively plansible that the ground states corresponding to
"middle-level" surfaces will be "dominant". This observation was
made formal in Ref's for the model on a cubic lattice.

\section{Model with an external field}

Let $M$ be the set of all unit balls with vertices in $V$, i.e.
$$M=\{\{x\}\bigcup S_1(x): \forall x\in V\}.$$

By the {\it restricted configuration} $\sigma_b$ we mean the restriction of a configuration $\sigma$ to a ball $b\in M$. Let $c_b$ denote the center of a unit
ball $b$. The energy of a configuration $\sigma_b$ on $b$ is defined by the formula

\begin{equation}\label{2}
U(\sigma_b)=-\frac{1}{2} J\sum_{\langle x, y \rangle, \atop x, y
\in b}|\sigma(x)-\sigma(y)|+\alpha_{c_b} \sigma(c_b).
\end{equation}

Note that $U(\sigma_b)$ has finitely many values for arbitrary
configuration $\sigma$.\\

\textbf{Definition 3.1.} \emph{A configuration $\varphi$ is called
a ground state for the Hamiltonian (\ref{1}), if
$$U(\varphi_b)=\min\{U_{\psi_b}\}$$
for any $\psi\in \Omega$ and $b\in M$.}\\

\textbf{Theorem 3.1.} \emph{Let $k=2$, $m=2$. For the SOS model
with arbitrary non-zero external field, if a translation-invariant
configuration is a ground state, then the external field is
translation-invariant.}

\textbf{Proof.} We shall prove that \ if  translation-invariant
configuration $\sigma(x)=2, \forall x\in V$ is a ground state,
then the external field is translation-invariant. 

Assume,
$\alpha_x\in\{\alpha_0, \alpha_1,...,\alpha_n,...\},\forall x\in
V$.

Let $\sigma(x)=2, \forall x\in V$ be a ground state. Then the energy of the unit balls $b\in M$ may be one of the following:
$$2\alpha_0,2\alpha_1, 2\alpha_2,...,2\alpha_n,... .$$

Since $\sigma(x)=2, \forall x\in V$ is a ground state the energy
$2\alpha_0$ must be minimal. From minimality of this energy for
variable of external field we get the following set
$\{(\alpha_0,\alpha_1,...,\alpha_n): 2\alpha_0\leq2\alpha_1,...,
2\alpha_0\leq2\alpha_n,...\}$. From minimality of $2\alpha_1$ we
take the set $\{(\alpha_0,
\alpha_1,...,\alpha_n):2\alpha_1\leq2\alpha_0,
2\alpha_1\leq2\alpha_2,..., 2\alpha_1\leq2\alpha_n,...\}$ etc.
Consequently we take the following:
$$\{(\alpha_0, \alpha_1,..., \alpha_n,...): \alpha_0\leq\alpha_1,..., \alpha_0\leq\alpha_n,...\}\cap$$
$$\cap \{(\alpha_0, \alpha_1,..., \alpha_n,...):\alpha_1\leq\alpha_0,
\alpha_1\leq\alpha_2,..., \alpha_1\leq\alpha_n,...\}\cap...$$
$$\cap\{(\alpha_0, \alpha_1,..., \alpha_n,...):\alpha_n\leq\alpha_0, \alpha_n\leq\alpha_1,...\}\equiv$$
$$\equiv\{(\alpha_0, \alpha_1,...,\alpha_n,...):\alpha_0=\alpha_1=...=\alpha_n...\},$$
i.e. external field must be a translation-invariant.

\textbf{This finishes the proof of Theorem 3.1.}

\begin{remark}
Note that  configuration $\sigma(x)=0, \forall x\in V$ is a
translation-invariant, but it may be ground states for any
$\alpha\geq0$ and the configuration $\sigma(x)=1, \forall x\in V$
is a ground state, if the external field is equal to zero.
\end{remark}

SOS model with a translation-invariant external field, i.e. $\alpha_{x}=\alpha$, $\forall x\in V$, is defined by the following Hamiltonian:
\begin{equation}\label{2}
H(\sigma)=-J\sum_{\langle x,y\rangle\in
L}|\sigma(x)-\sigma(y)|+\alpha\sum_{x\in V}\sigma(x),
\end{equation}
were  $J,\alpha \in \mathbb{R}.$ \   The energy of configuration
$\sigma_{b}$ on $b$ is defined by the formula
\begin{equation}\label{2}
U(\sigma_b)=-\frac{1}{2} J\sum_{\langle x, y \rangle, \atop x, y
\in b}|\sigma(x)-\sigma(y)|+\alpha \sigma(c_b).
\end{equation}
\   It is not difficult to prove the following Lemma.

\begin{lem} Let $k=2$. For each configuration $\sigma_b,$ we
have the following
$$U(\sigma_b)\in\{U_{1}(\sigma_{b}),....,U_{18}(\sigma_{b})\},$$ where
$$U_{1} (\sigma_{b})=-\frac{J}{2} ,\ \  \ \ \ \ \ U_{2} (\sigma_{b})=-J,\ \ \ \ \ \ U_{3} (\sigma_{b})=-\frac{3J}{2},$$
$$ U_{4} (\sigma_{b})=-2J ,\ \ \ \ \ \ U_{5} (\sigma_{b})=-\frac{5J}{2},\ \ \ \ \ U_{6} (\sigma_{b})=-3J,$$
$$ U_{7} (\sigma_{b})=-\frac{J}{2}+\alpha,\ \ U_{8}(\sigma_{b})=-J+\alpha,\ \ U_{9} (\sigma_{b})=-\frac{3J}{2}+\alpha ,$$
$$U_{10} (\sigma_{b})=-\frac{J}{2}+2\alpha,\ U_{11} (\sigma_{b})=-J+2\alpha,\ U_{12} (\sigma_{b})=-\frac{3J}{2}+2\alpha,$$
$$ U_{13} (\sigma_{b})=-2J+2\alpha,\ \ U_{14} (\sigma_{b})=-\frac{5J}{2}+2\alpha ,\ \ U_{15} (\sigma_{b})=-3J+2\alpha,$$
$$U_{16} (\sigma_{b})=\alpha,\ \ \ \ \ U_{17} (\sigma_{b})=2\alpha,\  \ \ \ \ \ U_{18} (\sigma_{b})=0. $$
\end{lem}

For every $i=\overline{1,18}$, we put
$A_{i}=\{(J,\alpha):U_{i}(\sigma_{b})\leq
U_{j},j=\overline{1,18}\}.$

\   Quite combersome but not difficult calculations show that:

$A_{1}=A_{2}=A_{3}=A_{4}=A_{5}=\{{(J,\alpha):J=0,\alpha\geq0}\},$

$A_{6}=\{(J,\alpha):J\geq0, \alpha\geq0\},\ \ \ A_{7}=A_{8}=A_{9}=\{(J,\alpha):J=0, \alpha=0\},$

$A_{10}=A_{11}=A_{12}=A_{13}=A_{14}=\{(J,\alpha):J=0,\alpha\leq0\},$

$A_{15}=\{(J,\alpha):J\geq0, \alpha\leq0\},\ \ \ A_{16}=\{(J,\alpha):J\leq0, \alpha=0\},$

$A_{17}=\{(J,\alpha):J\leq0, \alpha\leq0\},\ \ \ A_{18}=\{(J,\alpha):J\leq0, \alpha\geq0\}.$

The following theorem describes the necessary condition for a
configuration to be a ground state for the SOS model with
arbitrary non-zero external fields.

\textbf{Theorem 3.3.} \emph{For the SOS model with arbitrary
non-zero external field, if the external field is
translation-invariant, then arbitrary ground state is
translation-invariant.}

\textbf{Proof.} Let $\sigma$ be arbitrary ground state. For any
$b\in M$ we consider the following sets:
\begin{center}
$\Omega_{b,i}=\{\sigma_{b}:\sigma_{b}(c_{b})=i\}, \ i=0,1,2.$
\end{center}

May be the following cases:

 1) Let at least two sets $\Omega_{b,i}, \ i=0,1,2$  are nonempty:

  a) If the sets  $\Omega_{b,2}$ and $\Omega_{b,3}$  are nonempty, then \  $\sigma$ \  configuration is no translation-invariant and it is ground state on the set
  $ A_{i}\bigcap A_{j}=\{(J,\alpha):\alpha=0\},$

  $\ \ i\in\{7,8,9,16\},$
  $j\in\{10,11,12,13,14,15,17\}$, i.e external field must equal to zero.

  b) If the sets $\Omega_{b,1}$ and $\Omega_{b,2}$  are nonempty,\ then \ $\sigma$ \  configuration is no translation-invariant and it is ground state on the set
   $ A_{i}\bigcap A_{j}=\{(J,\alpha):\alpha=0\},$
   $  i\in\{1,2,3,4,5,6,18\},\ \ \ \  j\in\{7,8,9,16\}$, i.e external field must equal to zero.

  c) If sets $\Omega_{b,1}$ and $\Omega_{b,3}$  are nonempty, then\  $\sigma$\  configuration is no translation-invariant and it is ground state on the set
   $ A_{i}\bigcap A_{j}=\{(J,\alpha):\alpha=0\},\ i\in\{1,2,3,4,5,6,18\},$
   $j\in\{10,11,12,13,14,15,17\}$, i.e external field must equal to zero.

 2) If only one of the sets is nonempty and another two sets is empty, then $\sigma $ is translation-invariant.

\textbf{ This finishes the proof of Theorem 3.3.}

\begin{remark}
In [9] Ising model with non-zero external field is considered and
a necessary and sufficiency conditions for the external field are
found to make a translation-invariant configuration a ground state.
 Note that for model of SOS such necessary and sufficiency
conditions a not known.
\end{remark}

We let $GS(H)$ denote the set of all ground states of the
Hamiltonian $H$ {(see (3.2))}.

\textbf{Theorem 3.4.} \emph{For the SOS model with non-zero
translation-invariant external field (i.e. for the Hamiltonian
(3.2)).}

 a) If $(J,\alpha)\in A_{17}$ \mbox{then} $GS(H)=\{\{\sigma(x)=2,\forall x\in V\}\}.$

 b) If $(J,\alpha)\in A_{18}$ \mbox{then} $GS(H)=\{\{\sigma(x)=0,\forall x\in V\}\}.$

\textbf{Proof.} a) Consider the configuration $ \sigma(x)=2,
\forall x \in V .$ \ \ For any $ b\in M$ by $(3.3)$ we have
$U(\sigma_{b})=U_{17}.$  Thus the configuration
$\sigma(x)=2,\forall x\in V$ is ground state on the set $A_{17}.$

b) Consider the configuration $ \sigma(x)=0, \forall x \in V .$ \
\ For any $ b\in M$ by $(3.3)$ we have $U(\sigma_{b})=U_{18}.$
Thus the configuration $\sigma(x)=0,\forall x\in V$ is ground
state on the set $A_{18}.$

\textbf{This finishes the proof of Theorem 3.4.}

\begin{remark}
1) Note that if $\sigma(x)=1, \forall x\in V$  is a ground state,
then $(J,\alpha)\in A_{16},$ i.e. an external field is equal to
zero.

2) In \cite{3} periodic ground states for the Ising model with two
step interactions on the Cayley tree and with zero external fields
are described. In \cite{4} weakly periodic ground states for the
Ising model with competing interactions and with zero external
field are described.
\end{remark}

So, obviously seen from (Theorem  3.4), when an external field is
non-zero translation-invariant, all ground states for the SOS model are translation-invariant.

\section {Model with a periodic external field}

Let $G_k^{(2)}=\{x\in G_{k}:|x| \mbox { is\ even} \},$ where $ |x| $ means length of the word $x$. 
Now we shall study $G_k^{(2)}-$ periodic ground states for the SOS
model with $G_k^{(2)}-$periodic external field.

SOS model with $G_k^{(2)}-$periodic external field is defined
according to the following Hamiltonian:
\begin{equation}\label{5}
H(\sigma)=-J\sum_{\langle x,y\rangle\in
L}|\sigma(x)-\sigma(y)|+\sum_{x\in V}\alpha_x\sigma(x),
\end{equation}
where $J, \alpha_{x} \in \mathbb{R}$ and
$$
 \alpha_x=\left\{ \begin{array}{ll}
 \alpha_1, \, \mbox{if} \ \ x\in G_k^{(2)},\\[2mm]
 \alpha_2,\,  \mbox{if} \ \ x\in G_k\setminus G_k^{(2)},\\
      \end{array} \right.
$$
where $\alpha_1\neq \alpha_2$ and $G_k^{(2)}=\{x\in G_k: |x| \ \
\mbox{is even}\}$.

The energy of a configuration $\sigma_b$ on $b$ is defined by the formula
\begin{equation}\label{6}
U(\sigma_b)=-\frac{1}{2} J\sum_{x: \langle x,c_b\rangle \in
L}|\sigma(x)-\sigma(c_b)|+\alpha_{c_b} \sigma(c_b).
\end{equation}
It is not difficult to prove the following.

\begin{lem} We have $$U(\sigma_b)\in \{U_{1} (\sigma_{b}),
U_{2} (\sigma_{b}), U_{3} (\sigma_{b}), U_{4} (\sigma_{b}), ..., U_{29} (\sigma_{b})\}$$
 for all $\sigma_b$.\ Were
$$U_{1} (\sigma_{b})=-\frac{J}{2} ,\ \ \ \ \ U_{2} (\sigma_{b})=-J,\ \ \ \ U_{3} (\sigma_{b})=-\frac{3J}{2},\ \ \  \  U_{4} (\sigma_{b})=-2J,$$
$$U_{5} (\sigma_{b})=-\frac{5J}{2},\ \ \  \ \ \ \ \ \ \ \    U_{6} (\sigma_{b})=-3J,\ \ \ \ \  \ \ \ \ \ \ \    U_{7} (\sigma_{b})=-\frac{J}{2}+\alpha_{1},$$
$$U_{8} (\sigma_{b})=-J+\alpha_{1},\ \ \ \ \ U_{9} (\sigma_{b})=-\frac{3J}{2}+\alpha_{1} , \ \ \ \ U_{10} (\sigma_{b})=-\frac{J}{2}+2\alpha_{1},$$
$$U_{11} (\sigma_{b})=-J+2\alpha_{1},\ \  U_{12} (\sigma_{b})=-\frac{3J}{2}+2\alpha_{1},\ \  U_{13} (\sigma_{b})=-2J+2\alpha_{1},$$
$$U_{14} (\sigma_{b})=-\frac{5J}{2}+2\alpha_{1},\ \ \  \ \ \ \ U_{15} (\sigma_{b})=-3J+2\alpha_{1}\ \  \  \ \ \ \  U_{16} (\sigma_{b})=\alpha_{1},$$
$$U_{17} (\sigma_{b})=2\alpha_{1},\ \ \ \ \ \ \ \  \ \ \ \  \ \ \ U_{18} (\sigma_{b})=0,\ \ \ \ \ \ \  \ \ \\ \ \ \ U_{19} (\sigma_{b})=-\frac{J}{2}+\alpha_{2},$$
$$U_{20} (\sigma_{b})=-J+\alpha_{2},\ \ \ \  U_{21} (\sigma_{b})=-\frac{3J}{2}+\alpha_{2},\ \ \ \\ \ U_{22} (\sigma_{b})=-\frac{J}{2}+2\alpha_{2},$$
 $$U_{23} (\sigma_{b})=-J+2\alpha_{2},\ \ \ U_{24} (\sigma_{b})=-\frac{3J}{2}+2\alpha_{2},\  \ U_{25} (\sigma_{b})=-2J+2\alpha_{2},$$
$$U_{26} (\sigma_{b})=-\frac{5J}{2}+2\alpha_{2},\ \ \ \ \ \ U_{27} (\sigma_{b})=-3J+2\alpha_{2},\ \ \ \ \ \ \ \  U_{28} (\sigma_{b})=\alpha_{2},$$
$$U_{29} (\sigma_{b})=2\alpha_{2}.$$
 \end{lem}
\textbf{Definition 4.1.}  \emph{A configuration $\varphi$ is called a
ground state of the Hamiltonian (\ref{5}), if
$$U(\varphi_b)=\min \{U_{1} (\sigma_{b}),
U_{2} (\sigma_{b}), U_{3} (\sigma_{b}), ..., U_{29}
(\sigma_{b})\}$$ for all $b\in M$.}

For a fixed $m=1,2,3,....,29$, we set $$ A_{m}=\{(J, \alpha_0,
\alpha_1)\in \mathbb{R}^3 : U_{m}=\min \{U_{1} (\sigma_{b}),U_{2}
(\sigma_{b}), U_{3} (\sigma_{b}), ..., U_{29} (\sigma_{b})\}.$$

Quite cumbersome but not difficult calculations show that

$A_{1}=A_{2}=A_{3}=A_{4}=A_{5}=\{(J, \alpha_0, \alpha_1) \in \mathbb{R}^3 :J=0, \ \alpha_1\geq0,\  \alpha_2\geq0\},$

$A_{6}=\{(J, \alpha_0, \alpha_1) \in \mathbb{R}^3 :J\geq0, \alpha_1\geq0,\  \alpha_2\geq0\},$

$A_{7}=A_{8}=A_{9}=\{(J, \alpha_0, \alpha_1) \in \mathbb{R}^3 :J=0, \ \alpha_1=0,\  \alpha_2\geq0\},$

$A_{10}=...=A_{14}=\{(J, \alpha_0, \alpha_1) \in \mathbb{R}^3 :J=0, \ \alpha_1\leq0,\  \alpha_2\geq\alpha_1\},$

$A_{15}=\{(J, \alpha_0, \alpha_1) \in \mathbb{R}^3 :J\geq0, \ \alpha_1\leq0,\  \alpha_2\geq\alpha_{1}\},$

$A_{16}=\{(J, \alpha_0, \alpha_1) \in \mathbb{R}^3 :J\geq0, \ \alpha_1=0,\  \alpha_2\leq0\},$

$A_{17}=\{(J, \alpha_0, \alpha_1) \in \mathbb{R}^3 :J\geq0, \ \alpha_1\geq0,\  \alpha_2\leq\alpha_{1}\},$

$A_{18}=\{(J, \alpha_0, \alpha_1) \in \mathbb{R}^3 :J\leq0, \ \alpha_1\geq0,\  \alpha_2\geq0\},$

$A_{19}=A_{20}=A_{21}=\{(J, \alpha_0, \alpha_1) \in \mathbb{R}^3 :J=0, \ \alpha_1\geq0,\  \alpha_2=0\},$

$A_{22}=...=A_{26}=\{(J, \alpha_0, \alpha_1) \in \mathbb{R}^3 :J=0, \ \alpha_1\geq\alpha_{2},\  \alpha_2\leq0\},$

$A_{27}=\{(J, \alpha_0, \alpha_1) \in \mathbb{R}^3 :J\geq0, \ \alpha_1\geq\alpha_{2},\  \alpha_2\leq0\},$

$A_{28}=\{(J, \alpha_0, \alpha_1) \in \mathbb{R}^3 :J\geq0, \ \alpha_1\leq0,\  \alpha_2=0\},$

$A_{29}=\{(J, \alpha_0, \alpha_1) \in \mathbb{R}^3 :J\geq0, \
\alpha_1\leq\alpha_{2},\  \alpha_2\geq0\}.$ \\

\textbf{Theorem 4.1.}  \textbf{1)} \emph{The following
$G_2^{(2)}-$periodic configuration}
\begin{equation}\label{7}
 \sigma(x)=\left\{ \begin{array}{ll}
 0, \, \mbox{if} \, \ x\in G_2^{(2)},\\[2mm]
 1,\,  \mbox{if} \, \ x\in G_2\setminus G_2^{(2)},\\
 \end{array} \right.\end{equation}
\emph{is a ground state on set $(J,\alpha_0,\alpha_1)\in A_{3}\cap
A_{21}$ for the (\ref{5}) model};

\textbf{2)} \emph{The following $G_2^{(2)}-$periodic
configuration} \begin{equation}\label{7}
 \sigma(x)=\left\{ \begin{array}{ll}
 1, \, \mbox{if} \, \ x\in G_2^{(2)},\\[2mm]
 0,\,  \mbox{if} \, \ x\in G_2\setminus G_2^{(2)},\\
 \end{array} \right.\end{equation}
 \emph{is a ground
state on set  $(J,\alpha_0,\alpha_1)\in A_{3}\cap A_{9}$  for the
(\ref{5}) model};

\textbf{3)} \emph{The following $G_2^{(2)}-$periodic
configuration} \begin{equation}\label{7}
 \sigma(x)=\left\{ \begin{array}{ll}
 0, \, \mbox{if} \, \ x\in G_2^{(2)},\\[2mm]
 2,\,  \mbox{if} \, \ x\in G_2\setminus G_2^{(2)},\\
 \end{array} \right.\end{equation}
 \emph{is a ground
state on set  $(J,\alpha_0,\alpha_1)\in A_{6}\cap A_{27}$  for the
(\ref{5}) model};

\textbf{4)} \emph{The following $G_2^{(2)}-$periodic
configuration} \begin{equation}\label{7}
 \sigma(x)=\left\{ \begin{array}{ll}
 2, \, \mbox{if} \, \ x\in G_2^{(2)},\\[2mm]
 0,\,  \mbox{if} \, \ x\in G_2\setminus G_2^{(2)},\\
 \end{array} \right.\end{equation}
 \emph{is a ground
state on set  $(J,\alpha_0,\alpha_1)\in A_{15}\cap A_{6}$  for the
(\ref{5})
model}. \\

\textbf{Proof.} \textbf{1)} When we define the configuration for
the model of (\ref{5}) on the Cayley tree in the form of (4.3) then $\forall b\in M$ we have $\sigma(c_b)=0$ or $\sigma(c_b)=1$.

If $\sigma(c_b)=0$ then for $\forall x\in S_1(c_b)$ we have
$\sigma(x)=1$. In this case by (\ref{6}) we take
$U(\sigma_b)=U_{3}=\frac{-3}{2}J$.

If $\sigma(c_b)=1$ then $\sigma(x)=0$ for $\forall x\in S_1(c_b)$.
Then we have $U(\sigma_b)=U_{21}=\frac{-3}{2}J+\alpha_2.$

From these cases, $G_2^{(2)}-$periodic  configuration (see (4.3))
for the model of (\ref{5}) is ground state on the set of
$A_{3}\cap A_{21}=\{(J, \alpha_0, \alpha_1) \in \mathbb{R}^3 :J=0,
\alpha_1\geq0, \alpha_2= 0\}.$

\textbf{2)} If $\sigma(c_b)=1$ then for $\forall x\in S_1(c_b)$ we
have $\sigma(x)=0$. In this case by (\ref{6}) we take
$U(\sigma_b)=U_{9}=\frac{-3}{2}J+\alpha_{1}$.

If $\sigma(c_b)=0$ then $\sigma(x)=1$ for $\forall x\in S_1(c_b)$.
Then we have $U(\sigma_b)=U_{3}=\frac{-3}{2}J$.

From these cases, $G_2^{(2)}-$periodic  configuration (see (4.4))
for the model of (\ref{5}) is ground state on the set of
$A_{3}\cap A_{9}=\{(J, \alpha_0, \alpha_1) \in \mathbb{R}^3 :J=0,
\alpha_1=0, \alpha_2\geq 0\}.$

\textbf{3)} If $\sigma(c_b)=0$ then for $\forall x\in S_1(c_b)$ we
have $\sigma(x)=2$. In this case by (\ref{6}) we take
$U(\sigma_b)=U_{6}=-3J$.

If $\sigma(c_b)=2$ then $\sigma(x)=0$ for $\forall x\in S_1(c_b)$.
Then we have $U(\sigma_b)=U_{27}=-3J+2\alpha_2$.

From these cases, $G_2^{(2)}-$periodic  configuration (see (4.5))
for the model of (\ref{5}) is ground state on the set of
$A_{6}\cap A_{27}=\{(J, \alpha_0, \alpha_1) \in \mathbb{R}^3
:J\geq 0, \alpha_1\geq0, \alpha_2=0\}.$

\textbf{4)} If $\sigma(c_b)=2$ then for $\forall x\in S_1(c_b)$ we
have $\sigma(x)=0$. In this case by (\ref{6}) we take
$U(\sigma_b)=U_{15}=-3J+2\alpha_{1}$.

If $\sigma(c_b)=0$ then $\sigma(x)=2$ for $\forall x\in S_1(c_b)$.
Then we have $U(\sigma_b)=U_{6}=-3J$.

From these cases, $G_2^{(2)}-$periodic  configuration (see (4.6))
for the model of (\ref{5}) is ground state on the set of
$A_{15}\cap A_{6}=\{(J, \alpha_0, \alpha_1) \in \mathbb{R}^3
:J\geq0, \alpha_1=0, \alpha_2\geq0\}.$

\textbf{This finishes the proof of Theorem 4.1.}

\begin{remark}
The configurations

$\sigma'(x)=\left\{ \begin{array}{ll}
 1, \, \mbox{if} \, \ x\in G_2^{(2)},\\[2mm]
 2,\,  \mbox{if} \, \ x\in G_2\setminus G_2^{(2)}\\
 \end{array} \right.$\ \   and\ \ \  $ \sigma''(x)=\left\{ \begin{array}{ll}
 2, \, \mbox{if} \, \ x\in G_2^{(2)},\\[2mm]
 1,\,  \mbox{if} \, \ x\in G_2\setminus G_2^{(2)}\\
 \end{array} \right.$ 
 
  are ground states, if $ J=0, \alpha_{1}=0,
\alpha_{2}=0. $

 Note that in the case $ J=0, \alpha_{1}=0, \alpha_{2}=0 $   all configurations are ground states.
\end{remark}

\textbf{Acknowledgments.} The authors are grateful to Professor U.A.Rozikov for the useful discussions.

\begin{center}
\textbf{References}
\end{center}

{\small

\medskip

\medskip
{\small
\begin{tabular}{p{9cm}}
Rahmatullaev M.M.\\
Institute of mathematics, Tashkent, Uzbekistan;
Namangan State Universite, Namangan, Uzbekistan, e-mail: mrahmatullaev@rambler.ru\\
\end{tabular}
}

\medskip {\small
\begin{tabular}{p{9cm}}
Abdusalomova M. R.\\
Namangan State Universite, Namangan, Uzbekistan,  e-mail: mahliyo13abdusalomova@gmail.com\\
\end{tabular}
}

\medskip {\small
\begin{tabular}{p{9cm}}
Rasulova M.A.\\
The Ministry of International Affairs Namangan academic lyceum, Namangan, Uzbekistan,  e-mail: m$\_$rasulova$\_$a@rambler.ru\\
\end{tabular}
}

\label{lastpage}

\begin{enumerate}

\bibitem{1} Rozikov U.A. Gibbs measures on Cayley trees.World
scientific.2013.
\bibitem{Gan} N. N. Ganikhodzhaev, Group representation and automorphisms of the Cayley tree, Dokl. Akad. nauk
Resp. Uzbekistan, no. 4, 3 (1994) [in Russian].
 \bibitem{2}F.Mukhamedov, Ch.Hee Pah, M.Rahmatullaev,
H.Jamil. Periodic and Weakly Periodic Ground States for the
$\lambda-$ Model on Cayley Tree. 2017.Journal of Physics: Conf.
Series 949, 012021, doi:$10.1088/1742-6596/949/1/012021$.
\bibitem{karga} M. I. Kargapolov and Yu. I. Merzlyakov,Fundamentals of the Theory of Groups(Springer-Verlag, New
York-Heidelberg-Berlin, 1979). [Fundamentals of Group Theory(Nauka, Moscow, 1982)].
\bibitem{3}U.A.Rozikov. A contructite Description of Grond States and Gibbs Measures
     for Ising Model with two step interations on Cayley tree. 2006. Journal of statistical
     Physics. Vol. 122. N2, $217-235$.
\bibitem{4} M. M. Rahmatullaev. Description of Weakly Periodic Ground States of Ising Model with
Competing Interactions on Cayley Tree.2010.Applied Mathematics
$\&$ Information Sciences $4(2), 237-251$.
\bibitem{5} M.M.Rakhmatullaev, M.A.Rasulova. Periodic and Weakly Periodic Ground States for the Potts Model
with Competing Interactions on the Cayley Tree.2016. ISSN
$1055-1344$, Siberian Advances in Mathematics, Vol. 26, No.3,
pp.$215-½229$.
\bibitem{6} Rozikov U. A., Rahmatullaev M. M. Weakly Periodic Ground States and Gibbs Measures for the Ising
Model with Competing Interactions on the Cayley Tree, Theor. Math. Phys. \textbf{160}, No. 3, 1292--1300 (2009).
\bibitem{7} Rahmatullaev M. M., Rasulova M. A. Ground States for the Ising model with an external field on the Cayley tree, Uz. Math. Journal, No. 3, 147-155 (2018).
\bibitem{8} Rozikov U. A., Description of limiting Gibbs measures
for $\lambda-$models on the Bethe lattice. Sib.Math.J. 39(1998)
427-435.
\bibitem{9} Rozikov U. A., Describing uncountable number of Gibbs measures for
inhomogeneous Ising model, Theor. Math. Phys. 118 (1999) 95-104.
\bibitem{10} Ganikhodjaev N. N. and Rozikov U. A., Description of periodic
extreme Gibbs measures of some lattice models on the Cayley tree,
Theor. Math. Phys. 111 (1997) 480-486.
\bibitem{11} Ganikhodjaev N. N. and Rozikov U. A., On disordered phase in
the ferromagnetic Potts model on the Bethe lattice, Osaka J. Math.
37 (2000) 373-383.
\bibitem{12} Rozikov U. A., Suhov Y. M., Gibbs measures for SOS
models on a Cayley tree, Infinite Dimensional Analysis, Quantum
Probability and Related Topics, Vol. 9, No. 3 (2006) 471-488.
\bibitem{13} Mazel A. E. and Suhov Yu. M., Random surfaces with two-sided
constraints: An application of the theory of dominant ground
states, J. Statist. Phys. 64 (1991) 111-134.
\end{enumerate}
\end{document}